\documentclass[conference,letter]{IEEEtran}

\usepackage{mathrsfs}
\usepackage{amssymb}

\usepackage{cite}
\usepackage{amsfonts}
\usepackage{graphicx}
\usepackage{psfrag}
\usepackage{epsfig}
\usepackage{lettrine}
\usepackage{stfloats}
\usepackage{array}
\usepackage{diagbox}
\usepackage{enumerate}
\usepackage{amsthm}
\usepackage{amsmath}
\usepackage{cases}
\usepackage{booktabs}
\usepackage{multirow}
\usepackage{multicol}
\usepackage{algorithm}
\usepackage{algorithmic}
\usepackage{multirow}
\usepackage{amsmath}
\usepackage{xcolor}

\usepackage{graphics}
\usepackage{graphicx}
\usepackage{epsfig}

\begin{document}

\title{An Effective Limited Feedback Scheme for FD-MIMO Based on Noncoherent Detection and Kronecker Product Codebook }

\author{Lisi Jiang$^1$, Juling Zeng$^2$\\
$^1$State Key Laboratory of Networking and Switching Technology, Beijing University of Posts and Telecommunications\\
$^2$School of Computer and Information, China Three Gorges University\\
Email: julingzeng@ctgu.edu.cn}
\maketitle

\begin{abstract}
The low complexity quantization of channel state information (CSI) and the utilization of  vertical freedom of three dimension (3D) channels are two critical issues in the limited feedback design of  the \emph{full dimension multi-input-multi-output} (FD-MIMO) systems. In this paper, we propose an effective limited feedback scheme. We first employ Kronecker product based codebook (KPC) to explore the vertical freedom of 3D channels, extending the limited feedback from two dimension (2D) to 3D. Fruthermore, we use noncoherent sequence detection (NCSD) to quantify the CSI which includes both the vertical and horizontal channel information. This quantization method exploits the duality between codebook searching and NCSD to transform the CSI quntization on KPC to two parallel NCSD. The complexity is reduced  from  exponential to linear with the number of antennas. We also show the proposed scheme does not affect the diversity gain of the system by demonstrating the full diversity order. Monte Carlo simulation results show that  the proposed scheme provides at least 1.2dB coding gain compared with traditional 2D limited feedback schemes. Moreover, the proposed scheme outperforms other FD/3D CSI quantization schemes by 0.8dB coding gain with moderate complexity when the channel is highly spatially correlated.
\end{abstract}
\textbf{ Keywords: FD-MIMO, Limited feedback, CSI quantization, Kronecker product, Low complexity}

\section{Introduction}
\lettrine[lines=2]{A}{s} a key candidate technology for the fifth-generation (5G) mobile communications system, full dimension multi-input-multi-output (FD-MIMO) has attracted significant attention in the wireless industry and academia in the past few years\cite{YounsunKim2014Full}. Utilizing a large number of antennas in a two-dimension (2D) antenna array panel (AAS), FD-MIMO has two main advantages compared with traditional MIMO. On one hand, the 2D panel allows the extension of spatial separation, providing an extra vertical freedom to improve vertical coverage and overall system capacity\cite{dao20113d}. On the other hand, it supports up to at least 64 antennas and increases the range of transmission for improving power efficiency\cite{hoydis2013massive}.

The limited feedback system is critical in the realization of FD-MIMO. There are two key challenges need to be solved in the implementation of limited feedback. \emph{First}, 
 the current codebooks which are predominantly designed and optimized for 2D MIMO channels should be extended to 3D channels to entirely explore the freedom of vertical dimension of the 3D channels.
\emph{Second},  low complexity channel state information (CSI) quantization method should be designed for the limited feedback system to overcome the high quantization complexity caused by the large AAS in FD-MIMO.

To make use of the vertical freedom of 3D channels, Kronecker product based codebook (KPC) is usually employed in FD-MIMO systems\cite{xie2013limited,ying2014kronecker}. \cite{wang2014kronecker} demonstrated that the codewords distribution of KPC matches the distribution of optimal beamforming vector of 3D channels, showing the effectiveness of KPC.  \cite{xie2013limited} has adopted a KPC constructed by two Discrete Fourier Transform (DFT) codebooks which is easy to implement. However, the noncoherent sequence detection (NCSD) which can reduce the quantization complexity cannot be used on this DFT based KPC. \cite{ryan2009qam} has proposed a Phase-Shift Keying (PSK) codebook and used the NCSD to reduce quantization complexity. But the PSK codebook is designed for 2D channels. In this paper, we employ a codebook defined as the Kronecker product of two PSK codebooks, in which NCSD can be used by decomposing the codebook and channel. 

Conventional codebook searching based CSI quantization method feedback a binary index of the codeword chosen in a common vector quantized (VQ) codebook with the number of codewords exponential to the number of antennas, which leads to a exponential complexity of CSI quantization\cite{roh2006transmit}. In FD-MIMO systems where hundreds of antennas are deployed such complexity is too high to implement\cite{choi2013noncoherent}. \cite{roh2006transmit} proposed a tree searching based CSI quantization method to reduce complexity. However, this method needs a storage space exponential to the number of antennas, which is impossible in FD-MIMO. \cite{ryan2009qam} and \cite{choi2013noncoherent} has found  the duality between the problems of finding the optimal beamforming vector in the codebook and the noncoherent sequence detection (NCSD) which detects the channel vector with linear complexity. Utilizing the duality, CSI quantization can be transformed to NCSD where storage is not needed. However, such duality can only be utilized for PSK codebook but not for KPC.  By vestigating the characteristics of 3D channels and KPC, we first decomposite the channel vector into two sub-vectors representing horizontal and vertical channels respectively according to the channel phase information of 3D channels. Then, we transform the CSI quntization to two parallel NCSD by respectively using the duality on the two PSK codebooks which construct the KPC. The full diversity order is also demonstrated by us to show the proposed limited feedback does not affect the diversity gain of the system.

The remainder is organized as follows. Section \uppercase\expandafter{\romannumeral2} presents the
system model. Section \uppercase\expandafter{\romannumeral3} describes the construction of KPC.
Section \uppercase\expandafter{\romannumeral4}proposes the low complexity quantization method. Section \uppercase\expandafter{\romannumeral5} demonstrates the diversity order of the KPC . Simulation results are presented in Section \uppercase\expandafter{\romannumeral6} following conclusion in Section \uppercase\expandafter{\romannumeral7}.

\section{System Model}
\subsection{Beamforming Model with Feedback}
\begin{figure}[!htb]
\centering
\includegraphics[width=8cm]{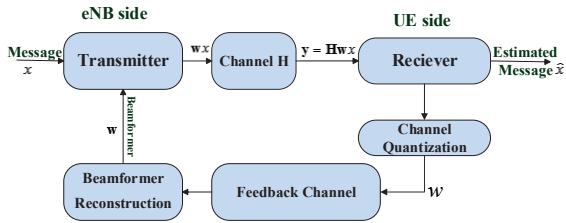}
\caption{MISO beamforming model with feedback.}
\label{feedback}
\end{figure}

We consider a multi-input single-output (MISO) communication system. The transmitter is equipped with $M_t$ transmit antennas and the receiver is equipped with 1 antenna. As shown in Fig. \ref{feedback}, for channel $ \textbf{H} \in \mathbb{C}^{M_t}$, the received signal $y \in \mathbb{C}$  can be written
as
\begin{equation}
\label{sysset}
y = \textbf{H}\textbf{w}x + n,
\end{equation}
where $\textbf{w} \in \mathbb{C}^{M_t}$ is the beamforming vector with $\left\| \textbf{w} \right\|_2^2 = 1$, $x \in \mathbb{C}$ is the message signal with $E[x] = 0$ and $E[{\left| x \right|^2}] = P$, and $n \in \mathbb{C}$ is additive complex Gaussian noise. For equal gain transmission (EGT), the beamforming vector has the property that $\left| {{w_t}} \right| = 1/\sqrt {{M_t}} $ for all $t$. Equal gain transmission is mainly considered in this paper because of its low peak to average power ratio (PAPR). We assume that $\textbf{H}$ is memoryless MIMO fading channels. The receiver sends the quantification of $\textbf{H}$ over a limited rate feedback channel. After receiving the feedback, the transmitter will construct a beamforming vector $\textbf{w}$ according to the quantified $\textbf{H}$. Aiming to focus on channel quantization design, we assumed that there are no channel estimation errors at the receiver or errors over the feedback channel, which means the perfect CSI.
\subsection{3D Channel Model}
\begin{figure}[!htb]
\centering
\includegraphics[width=8cm]{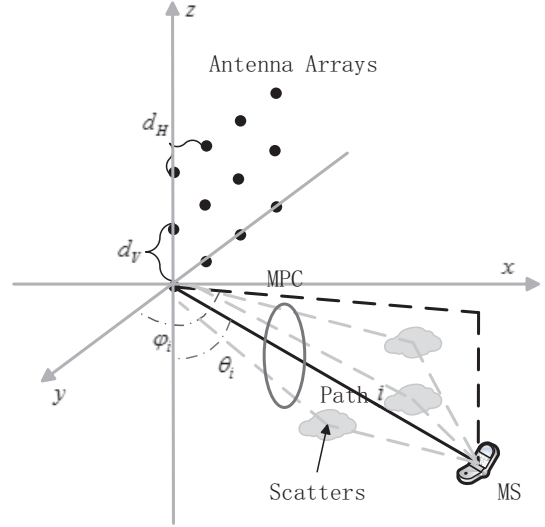}
\caption{3D Channel under UPA, $M_{tV}$ vertical antennas with $d_V$ wavelength spacing, and $M_{tH}$ horizontal antennas with $d_H$ wavelength spacing, ${M_{tH}}{M_{tV}} = {M_t}$. ${\varphi _i}$ and $\theta_i$ are the azimuth angle and the elevation angle for path $i$, respectively.}
\label{channelpicture}
\end{figure}

Uniformed Planar Arrays (UPA) is the most typical antenna array in FD-MIMO. As shown in Fig. \ref{channelpicture}, antennas are uniformly spaced across the planar. The departure and arrival angles are modeled by using the azimuth angle in X-Y plane and the elevation angle respected to the Z axis \cite{yong2005three} in 3D channel modeling. The 3D channel impulse response (CIR) $h(t)$ can be expressed as equation (\ref{ht1})
\begin{equation}
\label{ht1}
h(t) = \sum\limits_{i = 1}^{{I_{MPC}}} {{\alpha _i}(t)} \partial ({\Theta _i}),
\end{equation}
where ${\alpha _i}(t)$ is a zero-mean complex i.i.d. random variable, $\partial ({\Theta _i})$ is normalized impulse response to a single path with an Angle of Departure (AoD) ${\Theta _i}$ modeled in 3D. ${I_{MPC}}$ is the total number of multi-paths. The AoD of $i$-th multi-path component is given as ${\Theta _i} = {[{\theta _i},{\varphi _i}]^T}$. $0 \le {\varphi _i} \le 2\pi$ and $0 \le {\theta _i} \le \pi$ are the azimuth and elevation angles defined with respect to the positive y- and negative z-axis, respectively. Assuming that the antenna elements are vertically polarized, ${\Theta _i}$ only depends on ${\varphi _i}$ and ${\theta _i}$. The $\partial ({\Theta _i})$  for a size ${M_{tH}} \times {M_{tV}}$ UPA is given by
\begin{equation}
\partial ({\Theta _i}) = vec({\partial _H}(\mu ){\partial ^T}_V(\upsilon )),
\end{equation}
where ${\partial _H}(\mu)={[1,{e^{-j\mu }},...,{e^{-j({M_{tH}}-1)\mu}}]^T}$, $\mu=\frac{{2\pi}}{\lambda}{d_H}\cos({\varphi _i})\sin({\theta _i})$, ${\partial _V}(\upsilon)={[1,{e^{-j\upsilon}},...,{e^{-j({M_{tV}}-1)\upsilon}}]^T}$, $\upsilon=\frac{{2\pi}}{\lambda}{d_V}\cos({\theta _i})$. $\lambda$ is the wavelength of carriers and operator $vec()$ represents the operation of vectorization. ${d_H}$ and ${d_V}$ represent the distance between the antenna elements horizontal and vertical respectively. In common case, ${d_H}={d_V}=0.5\lambda$. The ${\Theta _i}$ of each AoD has a Gaussian distribution with a mean value of $\Theta$ and an angular spreading variance of $\sigma$.
Particularly, when the channel exhibits full spatial correlation, the channel model can be expressed as equation (\ref{fullcor}).
\begin{equation}
\label{fullcor}
h(t) = \alpha (t)\partial (\Theta),
\end{equation}
where $\alpha (t)$ is zero-mean complex i.i.d. random variable and $\partial (\Theta )$ is the NIR-SRP for AoD $\Theta$. The channel spatial correlation will increase as the number of multi-paths and the angular
spreading variance decrease. It worth noting that the channels are highly correlated in FD-MIMO systems, since the BSs deploys a large scale antenna array in a limited space. In addition, equation (\ref{fullcor}) is very important because the channel can be decomposed, which we will mention in section \uppercase\expandafter{\romannumeral4}.

\section{Kronecker product codebook construction}
To make use of the vertical freedom of 3D channels, Kronecker product based codebook (KPC) is usually employed in FD-MIMO systems\cite{xie2013limited,ying2014kronecker}. \cite{wang2014kronecker} demonstrated that the codewords distribution of KPC matches the distribution of optimal beamforming vector of 3D channels, showing the effectiveness of KPC.  \cite{xie2013limited} has adopted a KPC constructed by two Discrete Fourier Transform (DFT) codebooks which is easy to implement. However, the noncoherent sequence detection (NCSD) which can reduce the quantization complexity cannot be used on this DFT based KPC. \cite{ryan2009qam} has proposed a Phase-Shift Keying (PSK) codebook and used the NCSD to reduce quantization complexity. But the PSK codebook is designed for 2D channels. In this paper, we employ a codebook defined as the Kronecker product of two PSK codebooks, in which NCSD can be used by decomposing the codebook and channel. The Kronecker product based codebook $W_K$ can be generated as follows:
\begin{equation}
\label{codebookgeneration}
{W_K} = {W_V} \otimes {W_H},
\end{equation}
where operator $\otimes$ represents the Kronecker product. $W_H$ and $W_V$ represent the horizontal and vertical traditional PSK codebook respectively. We assume $M_{tV}$ and $M_{tH}$ represent the number of vertical antennas and horizontal antennas in UPA, respectively. $W_V$ is a PSK codebook with all possible sequences of PSK symbols with a constellation size of $N_V$. The length of $W_V$ is $M_{tV}$. Similarly, $W_H$ is allpossible sequences of PSK symbols of length $M_{tH}$ with constellation size $N_H$. The codebooks parameters are summarized in Table \ref{codebookpara}.
\begin{table}[h]
\centering
\caption{parameters of codebooks}
\label{codebookpara}
\begin{tabular}{@{}c|c|c@{}}
\toprule
\textit{\textbf{Codebook name}} & \textit{\textbf{Codewords vector length}} & \textit{\textbf{Constellation size}} \\ \midrule
$W_H$                      & $M_{tH}$                           & $N_H$                           \\
$W_V$                      & $M_{tV}$                           & $N_V$                           \\
$W_K$                      & $M_t=M_{tH}M_{tV}$                 & \emph{Not PSK Symbols}                      \\ \bottomrule
\end{tabular}
\end{table}

We assume that $w_V^1 =1/{M_{tV}}$ and $w_H^1 =1/{M_{tH}}$.




\section{Channel quantization scheme}
This section shows how the optimal codebook search over the proposed PSK-KPC can achieved with linear complexity.
\subsection{Duality Between Codebook Searching and Non-coherent Sequence Detection}
For codebook searching, to maximize the SNR, the receiver chooses the beamforming vector from the codebook according to
\begin{equation}
\label{searching}
\textbf{w}_{opt} = \arg \mathop {\max }\limits_{\textbf{v} \in C} \frac{{{{\left\| {\textbf{Hv}} \right\|}^2}}}{{{{\left\| \textbf{v} \right\|}^2}}},
\end{equation}
where $\textbf{H}$ denotes for the channel vector in Section \uppercase\expandafter{\romannumeral2}, $C$ is the codebook, and $\textbf{v}$ represents the codewords.

Then, for NCSD, we consider a single antenna noncoherent, block fading, additive white Gaussian noise (AWGN) channel. The received vector can be expressed as
\begin{equation}
\textbf{y} = \beta \textbf{x} + n,
\end{equation}
where $\textbf{x} \in \mathbb{C}^N$ is a vector of $N$ transmitted symbols, $n \in \mathbb{C}^N$ is a vector of i.i.d. AWGN, $\beta \in \mathbb{C}$ is an unknown deterministic channel which is assumed constant for a period of $N$ symbols. $\textbf{y} \in \mathbb{C}^N$ is the received signal. According to \cite{choi2013noncoherent}, the GLRT-optimal data estimate $\hat{\textbf{x}}^{GLRT}$ is by solving 
\begin{equation}
\label{detection}
\begin{split}
 &{\hat{\textbf{x}}^{GLRT}} = \arg \mathop {\min }\limits_{\hat{\textbf{x}} \in \mathbb{C}^N} \mathop {\min }\limits_{\hat{\beta} }{\left\| {\textbf{y} - \hat{\beta}\hat{\textbf{x}}} \right\|^2} \\
 &= \arg \mathop {\min}\limits_{\hat{\textbf{x}} \in \mathbb{C}^N} \mathop{\min}\limits_{\alpha \in {\mathbb{R}^ + }}\mathop{\min}\limits_{\theta \in [0,2\pi )} {\left\|\textbf{y}\right\|^2}+\alpha^2{\left\|\hat{\textbf{x}}\right\|^2}-2\alpha{\mathop{\rm Re}\nolimits} ({e^{j\theta }}{\textbf{y}^H}\hat{\textbf{x}})\\
 &= \arg \mathop {\min}\limits_{\hat{\textbf{x}} \in \mathbb{C}^N}\mathop{\min}\limits_{\alpha \in {\mathbb{R}^ + }}{\left\|\textbf{y}\right\|^2}+\alpha^2{\left\|\hat{\textbf{x}}\right\|^2}-2\alpha\left| {{\textbf{y}^H}\textbf{x}} \right|\\
 &=\arg \mathop {\max}\limits_{\hat{\textbf{x}} \in \mathbb{C}^N}\frac{{{{\left| {{\textbf{y}^H}\textbf{x}} \right|}^2}}}{{{{\left\| \textbf{x} \right\|}^2}}},
\end{split}
\end{equation}
where $\beta=\alpha e^{j\theta} $, $\alpha \in R^+$ and $\theta \in [0,2\pi)$.

Noting that in our MISO system ${\left\|\textbf{H}\textbf{v}\right\|}^2={{\left|\textbf{H}\textbf{v}\right|}^2}$, it can be easily checked from (\ref{searching}) and (\ref{detection}) that finding the optimal codeword and the NCSD problems are equivalent, i.e.,
\begin{equation}
\label{equivalence}
\textbf{w}_{opt}=\textbf{v}^{GLRT}=\arg\mathop {\min}\limits_{\textbf{v} \in C} \mathop {\min}\limits_{\beta}{\left\| \textbf{H}^T - \beta\textbf{v} \right\|^2},
\end{equation}
where $C$ denotes the codebook. Therefore, we can find the $\textbf{w}_{opt}$ using noncoherent block demodulator.
\subsection{Low Complexity Quantization for CSI}
Based on the equivalence between codebook searching and NCSD, low complexity algorithms can be used for CSI quantization. Maximum likelihood (ML) noncoherent PSK sequence detection can be performed using an algorithm in \cite{sweldens2001fast}. For $N$ symbols, the complexity is $O(N\log N)$. However, the channel vector to be quantified contains both horizontal and vertical channel phase information. Sequences of only PSK symbols cannot precisely present the 3D characteristic of the channel vectors. Therefore, noncoherent PSK sequence detection cannot be directly applied for KPC CSI quantization. Nevertheless, if we can separate the vertical channel phase information and the horizontal information to construct two sub-channel vectors, the noncoherent PSK sequence detection can then be applied to quantify the two sub-channel vectors. And the final quantization vector can be the Kronecker product of the two sub-channel quantization vectors.

\emph{\textbf {Step 1: Decomposition of the channel vector.}} Investigating the structure of channel vector in Section \uppercase\expandafter{\romannumeral2}, we find that some specific elements only contain the vertical channel phase information. Therefore, we can decompose the channel vector $\textbf{H}$ by 
\begin{equation}
\label{decomH}
{\hat{\textbf{H}}_V} = \{ \left. {\hat{h}_V^n} \right|\hat{h}_V^n = {h^{n{M_{tH}}}},n = 0,...,{M_{tV}-1}\},
\end{equation}
\begin{equation}
\label{decomV}
{\hat{\textbf{H}}_H} = \{ \left. {\hat{h}_H^m} \right|\hat{h}_H^m = {h^m},m = 0,...,{M_{tH}-1}\},
\end{equation}
where $h^k$ denotes the $k$-th element of vector $\textbf{H}$, $\hat{h}_H^m$ denotes the $m$-th element of vector $\hat{\textbf{H}}_H$ and $\hat{h}_V^n$ denotes the $n$-th element of vector $\hat{\textbf{H}}_V$. $M_{tH}$ and $M_{tV}$ are the number of antennas in the UPA horizontally and vertically, respectively. We denote the first element of a vector to be $0$-th.

After decomposition, $\hat{\textbf{H}}_V$ only contains the vertical channel information. $\hat{\textbf{H}}_H$ contains the least vertical channel information, which make itself present most horizontal channel information. In particular, when the channel is fully correlated, the NIR-SRP of the channel is a Kronecker product of two PSK sequences. Therefore, after decomposition, $\hat{\textbf{H}}_V$ and $\hat{\textbf{H}}_H$ are two PSK sequences, which completely represent the vertical channel information and horizontal channel information respectively.

\emph{\textbf {Step 2: Quantization using noncoherent PSK sequence detection.}} Once $\hat{\textbf{H}}_H$ and $\hat{\textbf{H}}_V$ are obtained, we can use the noncoherent PSK sequence detection to quantify them. In this case, the cost function (\ref{equivalence}) has the form as follows
\begin{equation}
\label{nondetH}
\textbf{w}^{opt}_H=\arg\mathop {\min}\limits_{\hat{\textbf{v}}_H \in W_H} \mathop {\min}\limits_{\theta \in [0,2\pi)}{\left\| \hat{\textbf{H}}_H^T - e^{j\theta}\hat{\textbf{v}}_H \right\|^2},
\end{equation}
\begin{equation}
\label{nondetV}
\textbf{w}^{opt}_V=\arg\mathop {\min}\limits_{\hat{\textbf{v}}_V \in W_V} \mathop {\min}\limits_{\theta \in [0,2\pi)}{\left\| \hat{\textbf{H}}_V^T - e^{j\theta}\hat{\textbf{v}}_V \right\|^2},
\end{equation}
where $W_H$ and $W_V$ refer to the horizontal and vertical PSK codebooks in Section \uppercase\expandafter{\romannumeral3}, respectively. $\hat{\textbf{v}}_H$ and $\hat{\textbf{v}}_V$ denote the codewords in $W_H$ and $W_V$. They are sequences of PSK symbols.

(\ref{nondetH}) and (\ref{nondetV}) reduce the search space from $\mathcal{R}=\mathbb{C}$ to $\mathcal{R}=e^{j[0,2\pi)}$ due to constant-modulus property of the codewords. The magnitude of $\beta$ does not influence the corresponding best codeword estimate. A corresponding best codeword estimate exists for each phase $\theta$. Therefore $[0,2\pi)$ can be partitioned into intervals correspond to obtaining the same codeword estimate. Based on the rotational symmetry of PSK symbols, the search space can be further reduced to $[0, 2\pi/M)$. $M$ denotes the constellation size. In our case, the search space of $\textbf{w}^{opt}_H$ can be reduced to $[0, 2\pi/N_H)$ and $\textbf{w}^{opt}_V$ can be reduced to $[0, 2\pi/N_V)$.

The algorithm in \cite{sweldens2001fast} first effectively calculates the cross over angles, indicating the nearest neighbor boundaries of $[0,2\pi/M)$. The cross over angle is defined as the value for $\theta$ in $[0, 2\pi/N_H)$ or $[0, 2\pi/N_H)$ where the cross over happens. The definition of crossing over can refer to \cite{sweldens2001fast}. Then the algorithm sorts the angels in order of phase, making the codeword estimates and corresponding angular metric update in a recursive manner. The final results are picked with the largest noncoherent likelihood. The algorithm is described in Algorithm 1. Note that the $g_H$ and $g_V$ in the Algorithm 1 are integral vectors whose elements are integer respectively taking value in $[0, N_{H}-1)$ and $[0, N_{V}-1)$ and their length are $M_{tH}$ and $M_{tV}$, respectively.
\begin{algorithm}[htb]
\caption{Low Complexity Quantization}\label{ag:quantization}
\begin{algorithmic}[1]
\REQUIRE ~~\\
$\hat{\textbf{H}}_H$ and $N_{H}$;\\
$\hat{\textbf{H}}_V$ and $N_{V}$;\\
\STATE $e_H=\exp(2\pi j/N_{H})$ and $e_V=\exp(2\pi j/N_{V})$;\\
\STATE Compute the phases of $\hat{\textbf{H}}_H$ and $\hat{\textbf{H}}_H$ in multiples of $2\pi/N_{H}$ and $2\pi/N_{V}$ by setting \\
$arg_H=\arg(\hat{\textbf{H}}_H)*N_{H}/2\pi$\\
$arg_V=\arg(\hat{\textbf{H}}_V)*N_{V}/2\pi$;\\
\STATE Round $arg_H$ and $arg_V$ and the results are $g_H$ and $g_V$;\\
\STATE Sort the cross over angles by sorting $g_H-arg_H$ and $g_V-arg_V$. The results are $u_H$ and $u_V$;\\
\STATE Compute the terms of the first inner product\\
$p_H=\overline{\hat{\textbf{H}}_H}.*e_H^{g_H} $ \\
$p_V=\overline{\hat{\textbf{H}}_V}.*e_V^{g_V} $;\\
\STATE Arrange all the terms of the recursion in the vectors.\\
$v_H=[\sum p_H; p_H(u_H)*(e_H-1)]$\\
$v_V=[\sum p_V; p_V(u_V)*(e_V-1)]$;\\
\STATE Compute all the inner products, take their absolute values and keep the index of the largest one as $b_H$ and $b_V$;\\
\STATE Build the best test vector\\
$ g_H(u_H(1 : b_H-1))=g_H(u_H(1: b_H-1))+1$\\
$ g_V(u_V(1 : b_V-1))=g_V(u_V(1: b_V-1))+1$;\\
\STATE Normalize $g_H$ and $g_V$ to have first component zero.
\STATE $\textbf{w}^{opt}_H=e_H^{g_H}$ and $\textbf{w}^{opt}_V=e_V^{g_V}$;\\
\ENSURE~~\\
$\textbf{w}^{opt}_H$;\\
$\textbf{w}^{opt}_V$;\\
\end{algorithmic}
\end{algorithm}

\emph{\textbf {Step 3: Final beamforming vector generation.}} After \emph{Step 2}, $\textbf{w}^{opt}_H$ and $\textbf{w}^{opt}_V$ are obtained. We can then construct the final beamforming vector
\begin{equation}
\textbf{w}^{opt}=vec({\textbf{w}^{opt}_H(\textbf{w}^{opt}_V)^T)}.
\end{equation}
$\textbf{w}^{opt}$ belongs to the KPC $W_K$.

\begin{figure}[!htb]
\centering
\includegraphics[width=8cm]{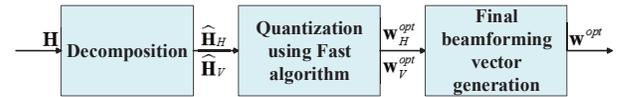}
\caption{CSI quantization of the KPC CSI.}
\label{quantization}
\end{figure}

Fig. \ref{quantization} presents the procedure of the KPC CSI quantization.

\subsection{Complexity}
In the quantization procedure, \emph{Step 1} and \emph{Step 3} have complexity $O(M_t)$. The complexity of \emph{Step 2} is mainly due to the noncoherent PSK sequence detection. The complexity of the algorithm is dominated by the sorting operation of cross over angels. The number of cross over angels equals to the length of the vector waiting to be detected. In our case, we use the noncoherent PSK sequence detection in parallel to quantify $\textbf{w}^{opt}_H$ and $\textbf{w}^{opt}_H$. Their length are $M_{tH}$ and $M_{tV}$. According to \cite{sweldens2001fast}, for $\textbf{w}^{opt}_H$ the complexity is $O(M_{tH}\log{M_{tH}})$ and for $\textbf{w}^{opt}_V$ the complexity is $O(M_{tV}\log{M_{tV}})$. Therefore, the complexity of \emph{Step 2} is $ O({M_{tH}}\log {M_{tH}})+O({M_{tV}}\log {M_{tV}})$. If ${M_{tH}} > \log {M_{tV}}$ and ${M_{tV}} > \log {M_{tH}}$, the overall complexity of channel quantization is $O(M_t)$.

Compared with traditional exponentially search complexity, our quantization complexity is linear to the number of antennas. However, it is worth noting that we obtain reduced complexity by decomposing the channel vector. This will certainly bring the performance loss. We will evaluate the performance loss in the Section \uppercase\expandafter{\romannumeral7}.

\section{Diversity order of the system}
Diversity order is an indicator of the performance of MIMO systems. The larger the diversity order is, the bigger the diversity gain we can obtain. For limited feedback systems, the diversity order of the system may be affected by the codebook design. To show that the proposed limited feedback scheme does not affect the diversity gain of the system, we demonstrate that the full diversity can be achieved.

Diversity order $D$ is achieved if the probability of symbol error, $P_e$, averaged over $H$ satisfies
\begin{equation}
\label{diversityorder}
\mathop {\lim }\limits_{{E_x}/{N_0} \to \infty } \frac{{\log {P_e}({E_x}/{N_0})}}{{\log ({E_x}/{N_0})}} =  - D,
\end{equation}
where $E_x$ is the signal power and $N_0$ is the noise power. For our $M_t \times 1$ system, the maximum achievable diversity order is $D = M_t$ \cite{lo1999maximum}. From them \textbf{Lemma 1} in \cite{love2003equal}, we know that in MISO system if the number of orthogonal vectors contained by the beamforming feasible set equals to the number of transmit antennas, the full diversity order can be achieved. Then, we have the corollary below.

\newtheorem{corollary}{\emph{Corollary}}
\begin{corollary}
The full diversity order is achieved when a MISO wireless system employes KPC.
\end{corollary}
\newtheorem{Proof}{\emph{Proof}}
\begin{proof}
Note that the set of all possible beamforming vectors are denoted as the beamforming feasible set and the set of all possible combining vectors are denoted as the combining feasible set in \cite{love2003equal}. In our case, we need to prove the beamforming feasible set contains $M_t$ orthogonal vectors. Because the length of the vectors in the beamforming feasible set is $M_t$, the largest number of orthogonal vectors equals to $M_t$. Therefore, if we demonstrate that a subset of the beamforming feasible set contains $M_t$ orthogonal vectors, which means the orthogonal vectors are no less than $M_t$, we can prove that the whole beamforming feasible set contains $M_t$ orthogonal vectors.
Then, we construct a subset of the beamforming feasible set by $U_K=U_H \otimes U_V $, where $U_H$ and $U_V$ are two Discrete Fourier Transform (DFT) square matrix presented below defined in \cite{wang2014kronecker}

$U_H$ and $U_V$ are easily to be proved unitary, then $U_K$ can be proved unitary as follows.
\begin{equation}
\begin{split}
{U_K}U_K^H &= ({U_H} \otimes {U_V}){({U_H} \otimes {U_V})^H}\\
  &= ({U_H} \otimes {U_V})(U_H^H \otimes U_V^H)\\
  &= ({U_H}U_H^H) \otimes ({U_V}U_V^H)\\
  &= {I_H} \otimes {I_V} = {I_K}.
\end{split}
\end{equation}
$U_K$ obviously belongs to the beamforming feasible set, which means there are at least $M_t$ orthogonal vectors in the beamforming feasible set. Since the length of the vectors in the beamforming feasible set is $M_t$, the beamforming feasible set can at most contain $M_t$ orthogonal vectors. Therefore, the full diversity is achieved.

\end{proof}

\section{simulation results and discussions}
\subsection{Simulation Setup}
Quadrature phase shift keying (QPSK) modulation is adopted in simulations. The BER is estimated using at least 10000 iterations per SNR point, where $2^{14}$ QPSK symbols are used for each iteration. Channels are generated according to (\ref{ht1}). The antenna spacing is usually set to be $0.5\lambda$ \cite{series2009guidelines}, where$\lambda$ is the wavelength of carriers. The spatial correlation $\rho$ is calculated according to \cite{yong2005three}, which is $0.66$ under antennas configuration in \cite{series2009guidelines}. $\rho$ is mainly related with the antenna spacing, the azimuth angle spread (AS) and elevation angle spread (ES). In common case, $\rho$ increases as the antenna spacing, the AS and the ES decrease.It is worth noting that since the BSs in FD-MIMO systems deploys a large scale antenna array within a limited space, the distance between antennas may be smaller than $0.5\lambda$, and the AS and the ES will also accordingly decrease \cite{yong2005three}. Therefore, $\rho$ will likely be higher than $0.66$ in future FD-MIMO systems.

We first compare the bit error rate (BER) performance of our proposed limited feedback scheme (3D-PSK) with traditional DFT codebooks based limited feedback scheme (2D-DFT) in two antenna configeration to show the efficiency of our scheme in the vertical freedom utilization. Then, we compare our 3D-PSK scheme with the 3D limited feedback scheme in \cite{wang2014kronecker} under four spatial correlation to show our 3D-PSK strike the balance between performance and complexity .

\subsection{3D-PSK vs. 2D-DFT}
\begin{figure}[!htb]
\centering
\includegraphics[width=8cm]{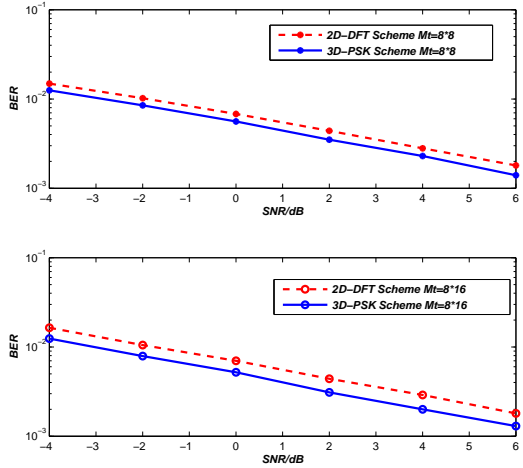}
\caption{BER versus SNR comparison with 2D-DFT when $\rho=0.66$.}
\label{2dcompare}
\end{figure}

Fig. \ref{2dcompare} shows that 3D-PSK outperforms the 2D-DFT by 0.8dB and 1.2dB coding gain respectively under both the two transmission configuration. This result shows that 3D-PSK makes full use of the vertical freedom of 3D channels, which will improve the system capacity. It can also be seen that  as the number of antennas increases, the coding gain gets larger,  which proves the superiority of 3D-PSK scheme in large scale FD-MIMO systems.

\subsection{3D-PSK vs. 3D-DFT}
\begin{figure}[!htb]
\centering
\includegraphics[width=8cm]{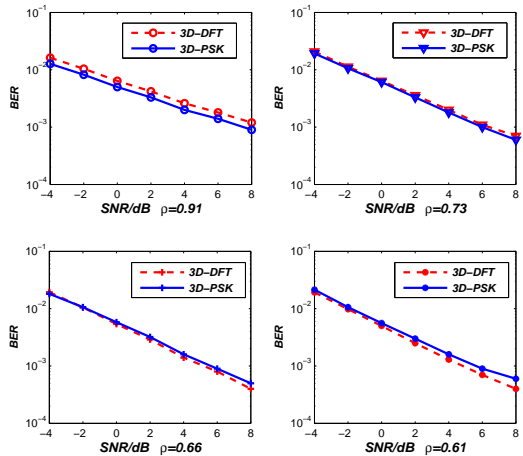}
\caption{BER versus SNR comparison between 3D-PSK and 3D-DFT when transmit antennas $M_t=8*8$.}
\label{3Dcompare}
\end{figure}

Fig. \ref{3Dcompare} compares 3D-PSK with 3D-DFT under  four spatial correlation: $\rho=0.91$, $\rho=0.73$, $\rho=0.69$ and $\rho=0.61$. When $\rho=0.91$, 3D-PSK earns about 1dB coding gain compared with the 3D-DFT. As the correlation decreases to $\rho=0.73$, the coding gain is down to about 0.4dB. When $\rho=0.66$, the two schemes perform nearly the same. When $\rho$ is down to $0.61$ 3D-DFT outperforms 3D-PSK . This validates our inference in Section \uppercase\expandafter{\romannumeral4} that our 3D-PSK is more appropriate for highly correlated systems, especially when $\rho$ is above $0.66$. SInce $\rho$ will likely be higher than $0.66$ in future FD-MIMO systems, 3D-PSK will have a very promising future.

Table \ref{my-label} further compars the complexity and BER between 3D-PSK and 3D-DFT under the four spatial correlation when SNR is set 2dB. We can see that 3D-PSK has a linear complaxity with better BER performance in highly correlated environment while 3D-DFT has a exponential complexity with better BER in less correlated environment. Since channels in FD-MIMO will be more and more correlated, our 3D-PSK will strike better banlance between complexity and performance than 3D-DFT.

\begin{table}[]
\centering
\caption{Complexity and BER comparison between 3D-PSK and 3D-DFT when transmit antennas $M_t=8*8$. }
\label{my-label}
\begin{tabular}{|c|c|c|c|c|}
\hline
$\rho$ & BER(3D-PSK)   & BER(3D-DFT)   & CPLX(3D-DFT) & CPLX(3D-PSK) \\ \hline
0.91   & $3.3*10^{-2}$ & $4.2*10^{-2}$ & $O(2^{M_t})$ & $O(M_t)$     \\ \hline
0.73   & $3.7*10^{-2}$ & $4.1*10^{-2}$ & $O(2^{M_t})$ & $O(M_t)$     \\ \hline
0.66   & $4.0*10^{-2}$ & $4.1*10^{-2}$ & $O(2^{M_t})$ & $O(M_t)$     \\ \hline
0.61   & $4.2*10^{-2}$ & $4.0*10^{-2}$ & $O(2^{M_t})$ & $O(M_t)$     \\ \hline
\end{tabular}
\end{table}

\section{conclusion}
In this paper, an effective limited feedback scheme for FD-MIMO has been proposed to explore the vertical freedom of 3D channels with moderate CSI quantiaztion complexity. We first employ KPC to extend the limited feedback from 2D to 3D. Then, we use NCSD to quantify the CSI which includes both the vertical and horizontal channel information and reduce the complexity from  exponential to linear with the number of antennas. We also show the proposed scheme does not affect the diversity gain of the system by demonstrating the full diversity order. Simulation results validate the efficiency of our quantization scheme and demonstrate the better performance of our scheme in  high correlated environments, making it more promising for FD-MIMO systems.
\section*{Acknowledgement}
This work is supported by National 863 Project (2014AA\\01A705), National Nature Science Foundation of China (6143\\1003,61421061).

{
\footnotesize
\bibliographystyle{IEEEtran}
\bibliography{mybibfile}
}
\end{document}